\documentclass[5p]{elsarticle}

\usepackage{lineno}
\usepackage[draft]{hyperref} 
\usepackage{amsmath}
\modulolinenumbers[5]

\journal{Physics Letters B}

\biboptions{numbers,sort&compress}
\newcommand{\tra}[4]{\ensuremath{{#1}_{#2}^{+} \rightarrow {#3}_{#4}^{+} }}

\begin{document}

\begin{frontmatter}

\title{Identification of significant $E0$ strength in the \tra{2}{2}{2}{1} transitions of $^{58, 60, 62}$Ni
}

\author[triumf,surrey]{L.J.~Evitts}
\author[triumf]{A.B.~Garnsworthy\corref{cor1}}
\ead{garns@triumf.ca}
\author[anu]{T.~Kib\'{e}di}
\author[triumf]{J.~Smallcombe}
\author[anu]{M.W.~Reed}
\author[nscl,msu]{B.A.~Brown}
\author[anu]{A.E.~Stuchbery}
\author[anu]{G.J.~Lane}
\author[anu]{T.K.~Eriksen}
\author[anu]{A.~Akber}
\author[anu,saudi]{B.~Alshahrani}
\author[anu]{M.~de Vries}
\author[anu]{M.S.M.~Gerathy}
\author[triumf]{J.D.~Holt}
\author[anu]{B.Q.~Lee\fnref{fn_oxford}}
\author[anu]{B.P.~McCormick}
\author[anu]{A.J.~Mitchell}
\author[triumf]{M.~Moukaddam\fnref{fn_surrey}}
\author[UK]{S.~Mukhopadhyay}
\author[anu]{N.~Palalani}
\author[anu]{T.~Palazzo}
\author[UK]{E.E.~Peters}
\author[UK]{A.P.D.~Ramirez}
\author[triumf]{S.R.~Stroberg\fnref{fn_reed}}
\author[anu]{T.~Tornyi}
\author[UK]{S.W.~Yates}

\cortext[cor1]{Corresponding author}
\fntext[fn_oxford]{Present address: Department of Physics, University of Oxford, Oxford, OX1 3PJ, United Kingdom}
\fntext[fn_surrey]{Present address: Department of Physics, University of Surrey, Guildford, Surrey, GU2 7XH, United Kingdom}
\fntext[fn_reed]{Present address: Physics Department, Reed College, 3203 SE Woodstock Blvd., Portland, OR 97202-8199, USA}

\address[triumf]{Physical Sciences Division, TRIUMF, 4004 Wesbrook Mall, Vancouver, B.C., V6T 2A3, Canada}
\address[surrey]{Department of Physics, University of Surrey, Guildford, Surrey, GU2 7XH, United Kingdom}
\address[anu]{Department of Nuclear Physics, Research School of Physics and Engineering, The Australian National University, Canberra, ACT 2601, Australia}
\address[saudi]{Department of Physics, King Khaled University, Abha, Kingdom of Saudi Arabia}
\address[nscl]{National Superconducting Cyclotron Laboratory, Michigan State University, East Lansing, Michigan 48824, USA}
\address[msu]{Department of Physics and Astronomy, Michigan State University, East Lansing, Michigan 48824, USA}
\address[UK]{Departments of Chemistry and Physics \& Astronomy, University of Kentucky, Lexington, Kentucky, 40506-0055, USA}

\begin{abstract}
The $E0$ transition strength in the \tra{2}{2}{2}{1} transitions of $^{58,60,62}$Ni have been determined for the first time following a series of measurements at the Australian National University (ANU) and the University of Kentucky (UK).
The CAESAR Compton-suppressed HPGe array and the Super-e solenoid at ANU were used to measure the $\delta(E2/M1)$ mixing ratio and internal conversion coefficient of each transition following inelastic proton scattering. Level half-lives, $\delta(E2/M1)$ mixing ratios and $\gamma$-ray branching ratios were measured at UK following inelastic neutron scattering.
The new spectroscopic information was used to determine the $E0$ strengths. These are the first $2^+ \rightarrow 2^+$ $E0$ transition strengths measured in nuclei with spherical ground states and the $E0$ component is found to be unexpectedly large; in fact, these are amongst the largest $E0$ transition strengths in medium and heavy nuclei reported to date.
\end{abstract}


\end{frontmatter}


\section{Introduction}
The electric monopole ($E0$) transition operator represents a change in the spatial distribution of the nucleons within the nucleus.
The strength of an $E0$ transition, $\rho^2 (E0)$, can be directly related to the difference in the mean-squared charge radii, $\langle r^2 \rangle$, and the degree of configuration mixing between the states involved. Thus $E0$ transitions are a sensitive probe for the interpretation of shape mixing and shape coexistence effects \cite{Heyde.83.1467,Wood1999}.

Despite their importance, the number of $E0$ transition strengths that have been measured experimentally is very limited \cite{Kibedi2005}. This deficiency is primarily due to the often complex nature of the required measurements and the necessity for electron spectroscopy which can be hindered by many sources of background. 
There is especially a lack of data for $E0$ transition strengths in $J\rightarrow J,J>0$ transitions and \tra{2}{}{2}{} cases have only been reported in deformed nuclei, mostly in the lanthanide region \cite{Wood1999}.

The stable nickel isotopes just above doubly magic $^{56}$Ni have been studied extensively with a number of spectroscopic probes and mechanisms to access excited states. Detailed muonic X-ray measurements \cite{Shera1976} and optical spectroscopy \cite{Steudel1980} indicate that the ground states are spherical with little variation between the isotopes. The spectroscopic quadrupole moment of the first 2$^+$ state in each of $^{58,60,62}$Ni is small \cite{Raghavan1989}, indicating that the first excited state is also close to spherical.

The previous experimental work in determining $\rho^2 (E0)$ values between 0$^+$ states has been performed in $^{58,60,62}$Ni \cite{Passoja1981}, and recently in $^{68}$Ni indicating the presence of shape coexistence at $N=40$ \cite{Suchyta.89.021301}. An investigation of $E2$ strengths has revealed the coexistence of oblate, spherical and prolate excitations in $^{66}$Ni below an excitation energy of 3\,MeV \cite{Leoni2017}.
However, measurements of the $E0$ strength in any \tra{2}{}{2}{} transition in the Ni isotopes have not been pursued previously.
Determining the $E0$ strength between 2$^+$ states requires the experimental determination of a number of quantities, namely, the half-life of the parent state, the branching ratio of the transition, the $\delta$($E2$/$M1$) multipole mixing ratio and the internal conversion coefficient.
In this Letter, we present the first experimental measurements of the $E0$ strength in the \tra{2}{2}{2}{1} transitions of $^{58, 60, 62}$Ni. These $\rho^2 (E0)$ values are the first determined in nuclei with spherical ground states and are unexpectedly some of the largest measured to date.

\section{Experimental details}
Measurements were performed at the Heavy Ion Accelerator Facility at the Australian National University (ANU), with proton beams of up to 9.2~MeV provided by the 14UD pelletron accelerator. 
Inelastic proton scattering was used to populate excited states in the nickel isotopes of interest by impinging the beam on isotopically enriched self-supporting targets of 1.4\,mg/cm$^2$ for $^{58}$Ni and 1.3\,mg/cm$^2$ for $^{60, 62}$Ni.

\begin{figure}[!htbp]
	\centering
	\includegraphics[width=1.0\linewidth]{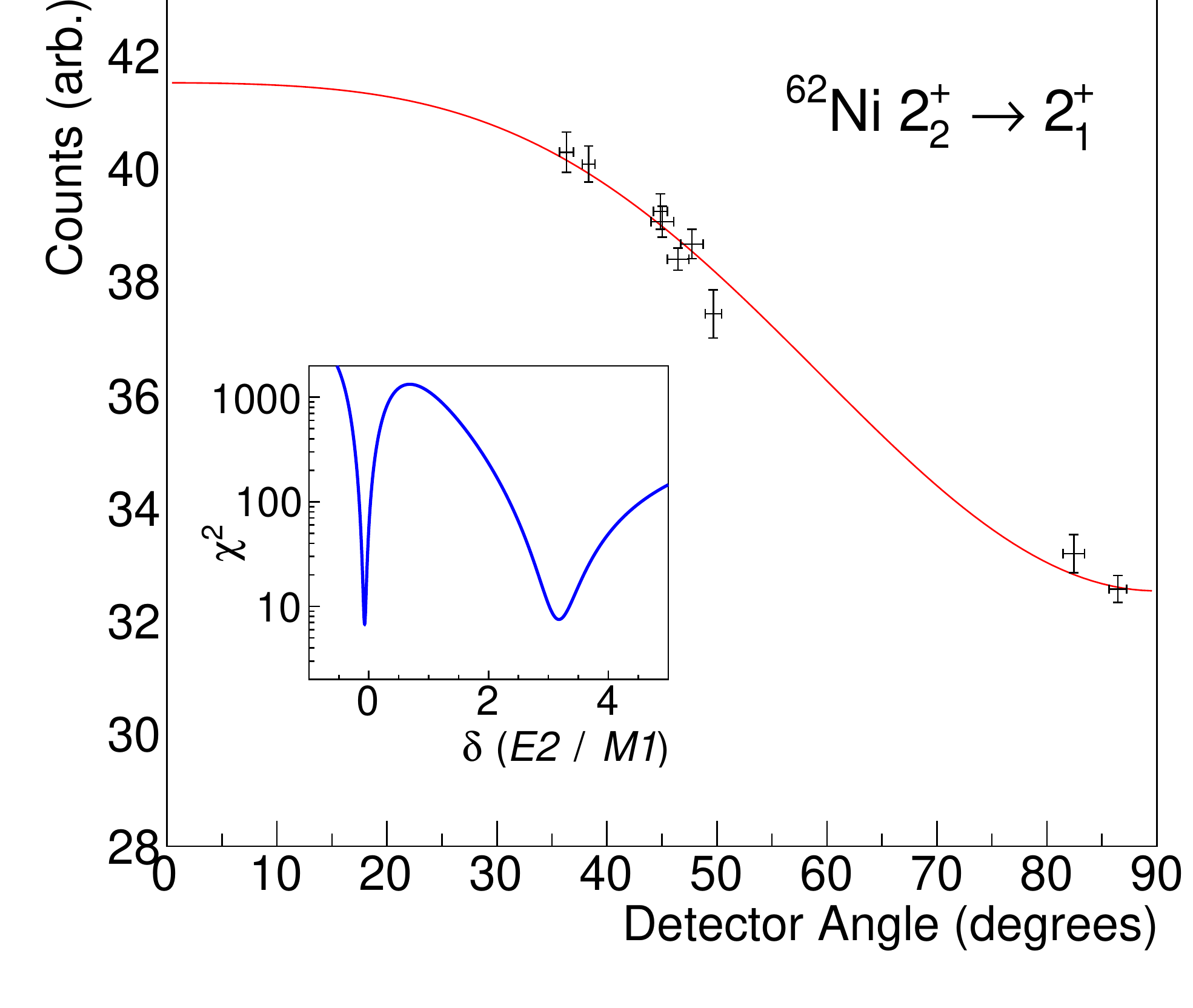}
	\caption{Example $\gamma$-ray angular distribution for the \tra{2}{2}{2}{1} transition in $^{62}$Ni from the ($p,p'\gamma$) measurement. The inset shows the associated $\chi^2$ minimization curve.}
	\label{fig:ang_dist}
\end{figure}

Angular distributions of $\gamma$ rays were measured using the CAESAR array composed of nine Compton-suppressed HPGe detectors. 
The angular distributions are sensitive to the $\delta(E2/M1)$ mixing ratio of a transition and the degree of angular momentum alignment provided to the nucleus in the reaction.
The degree of alignment of each of the 2$^+_2$ parent states of interest was determined from the angular distribution of the \tra{2}{2}{0}{1} transition of pure $E2$ multipolarity. This alignment factor was then used as a fixed parameter in determining the $\delta$ value of the \tra{2}{2}{2}{1} transition from a $\chi^2$ minimization analysis of the angular distribution with 7 degrees of freedom. An example for the \tra{2}{2}{2}{1} transition in $^{62}$Ni is shown in Fig. \ref{fig:ang_dist}.

Additional measurements were performed at the University of Kentucky (UK) Accelerator Laboratory. Angular distributions of $\gamma$ rays following the inelastic scattering of fast neutrons from natural nickel yielded branching ratios and $\delta(E2/M1)$ mixing ratios, for the \tra{2}{2}{2}{1} transitions, as well as the half-lives of the $2^+_2$ states in $^{58}$Ni and $^{60}$Ni from a Doppler-Shift Attenuation Method (DSAM) analysis. The details of the methods used are described in a previous study of $^{62}$Ni \cite{Chakraborty2011} that used an enriched $^{62}$Ni sample and from which we take the results for that isotope. The solid cylindrical scattering sample (45.94\,g of natural nickel metal; 68.08\% $^{58}$Ni, 26.22\% $^{60}$Ni) of 1.88\,cm diameter and 1.84\,cm height was bombarded with nearly monoenergetic neutrons ($\Delta E\approx$60\,keV) and $\gamma$ rays from the ($n,n'\gamma$) reaction were observed with a BGO Compton-suppressed HPGe detector.  Measurements at angles from 40$^\circ$ to 150$^\circ$ with respect to the incident beam were carried out at neutron energies of 2.42 and 2.90\,MeV.  In each case, the bombarding energy was chosen to yield significant population of the level of interest but to avoid feeding of the level from higher-lying states.  

\begin{figure}[!htbp]
	\centering
	\includegraphics[width=1.0\linewidth]{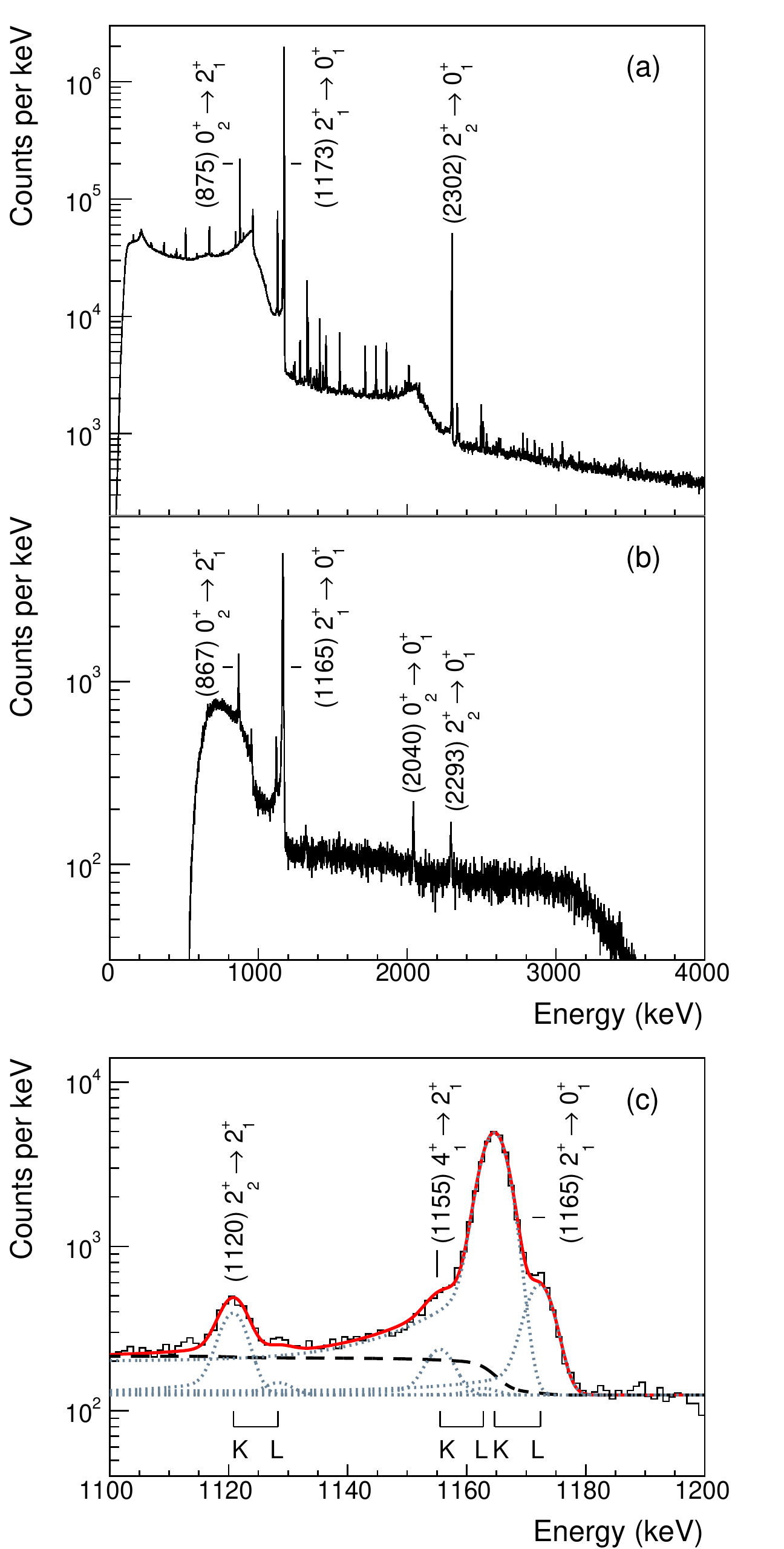}
	\caption{(a) $\gamma$-ray and (b) e$^-$ energy spectra for $^{62}$Ni from the Super-e spectrometer. A portion of the electron energy spectrum is shown in (c), demonstrating the procedure for fitting overlapping K and L electron peaks. Three pairs of K and L lines are indicated by the grey dashed lines on top of the black dashed background. The contribution from conversion in higher atomic orbitals is omitted as it is $<$1.5\% that of the K shell. The reduced $\chi^2$ value of the fit to the 1100-1200\,keV data is 1.2.
}
\label{fig:spectra}
\end{figure}

The superconducting electron spectrometer, Super-e \cite{Kibedi1990}, located at the ANU was used to measure internal conversion coefficients of the same transitions. Super-e consists of a super-conducting solenoid magnet, a set of six 9\,mm thick Si(Li) detectors chosen to be suitable for detecting electrons up to $\sim$3.5\,MeV and a single Compton-suppressed HPGe detector to allow for simultaneous measurements of electrons and $\gamma$ rays. Data were collected from both detectors in singles mode. Example spectra from the Super-e measurements using a $^{62}$Ni target are shown in Fig. \ref{fig:spectra} along with an example of the electron peak-fitting procedure.  
The electrons emitted from the target, tilted at 45$^{\circ}$ to the beam axis, were transported by a magnetic field oriented perpendicular to the beam axis. The magnetic field was selected so that electrons of a specific energy follow a helical path that transports them around two baffles and through a diaphragm to the Si(Li) detector array located 35\,cm from the target. The magnetic field was swept over a suitable range in small steps to cover the electron energies of interest. The measurement period at each magnetic field value was controlled by the current recorded in a Faraday cup downstream of the target so that the integrated beam current at each field value was the same.

The $\rho^2 (E0)$ value can be determined \cite{Kibedi2005} from 

\begin{equation}
\rho^2 (E0) = \frac{1}{\Omega_K (E0) \cdot \tau_K (E0)}
\end{equation}

\noindent where $\Omega_K$ is an electronic factor obtained from atomic theory \cite{Kibedi2008} and $\tau_K (E0)$ is the partial mean lifetime of the $E0$ transition converted in the atomic K shell.  The value of $\tau_K (E0)$ is calculated using the decay branch of the $E0$ component, $\lambda_{E0}$, relative to the sum of all available decay modes, $\sum_i \lambda_i$, from the parent state i.e.,

\begin{equation}
\label{eq:tau_from_lamb}
\tau_K (E0) = \frac{\sum_i \lambda_i}{\lambda_{E0_K}} \cdot \frac{T_{1/2}}{\textnormal{ln} (2)}
\end{equation}

\sloppy
\noindent where $T_{1/2}$ is the half-life of the parent state. The $\delta$($E2$/$M1$) mixing ratio is essential experimental information for the determination of $\tau_K (E0)$. The available experimental values, such as $\delta$($E2$/$M1$), internal conversion coefficients and $\gamma$-ray branching ratios, can be used to calculate each available $\lambda_i$ for the transition of interest.
The value of $\rho^2 (E0)$ is dimensionless and, as the magnitudes are typically on the order of 10$^{-4}$ to 10$^{-1}$, experimental $\rho^2 (E0)$ values are generally quoted in milliunits.

\section{Experimental results}

\begin{figure*}[!ht]
	\centering
	\includegraphics[width=1.0\linewidth]{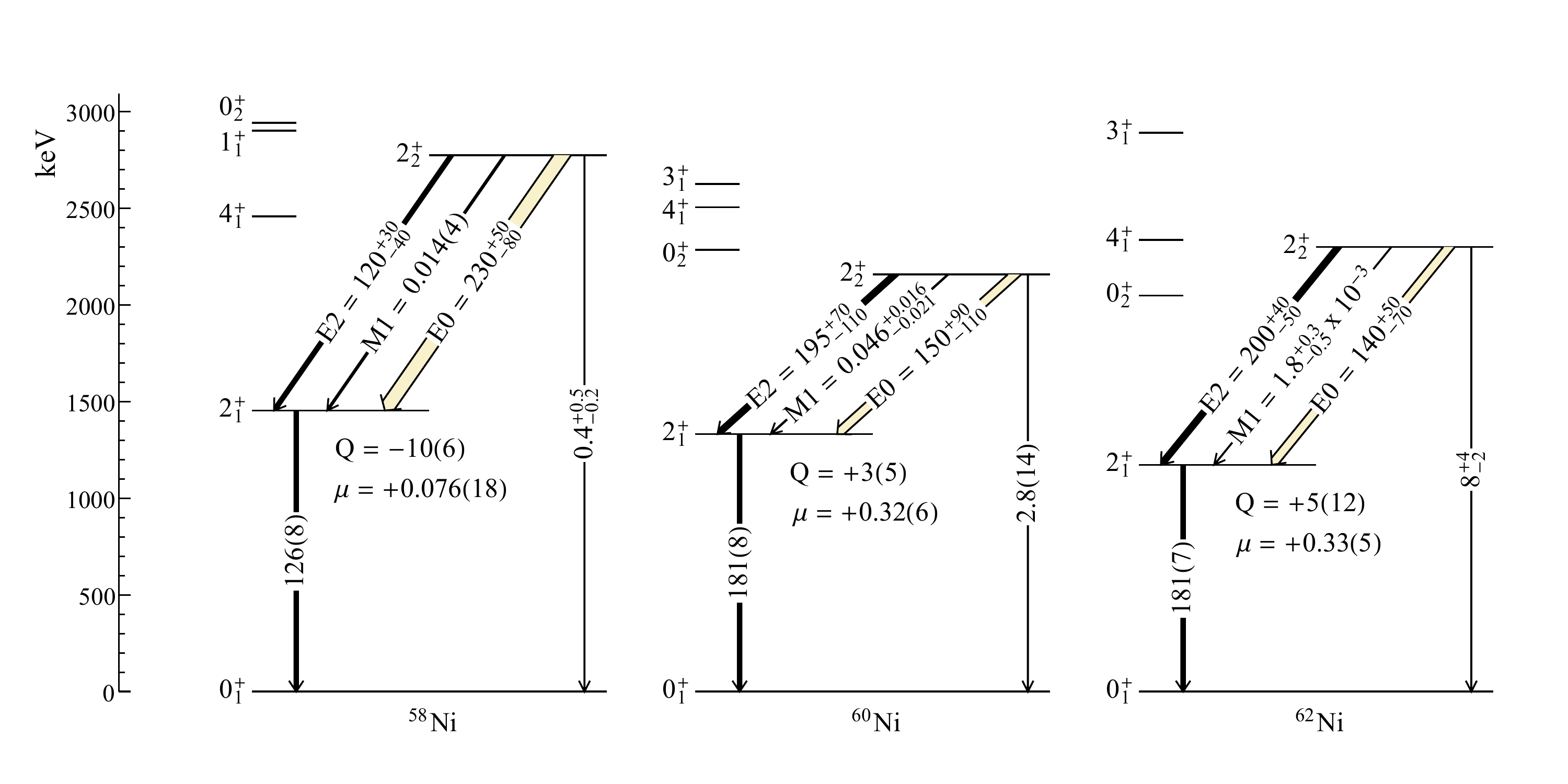}
	\caption{Partial level schemes showing low-lying excited states in $^{58,60,62}$Ni isotopes where the black filled arrows show B($M1$) [$\mu_N^2$] and B($E2$) [$e^2 fm^4$] values, while the yellow arrows show $\rho^2 (E0)\,\times\,10^3$ values \cite{Allmond.90.034309,Kenn.63.021302,NucData58,NucData60,NucData62}.  The magnetic moments \cite{Kenn.63.021302}, $\mu$, are in units of nuclear magnetons ($\mu_{N}$) and the spectroscopic quadrupole moments \cite{NucData58,NucData60,NucData62}, $Q$, are in units of $efm^2$.}
	\label{fig:summary_results}
\end{figure*}

The new experimental data, combined with the recent study of Chakraborty {\it et al.} \cite{Chakraborty2011}, have provided new values for the half-lives ($T_{1/2}$) of the 2$^+_2$ levels from a DSAM analysis as well as branching ratios (BR), $\delta (E2/M1)$ mixing ratios, and K-shell internal conversion coefficients ($\alpha_K$) of the \tra{2}{2}{2}{1} transitions in $^{58,60,62}$Ni. The branching ratios for $^{58,60}$Ni are from the ($n,n'\gamma$) measurements and are consistent with, but more precise than, those previously reported \cite{NucData58,NucData60}. The $T_{1/2}$ and branching ratio for $^{62}$Ni are taken from the previous UK ($n,n'\gamma$) study of Ref. \cite{Chakraborty2011}. All results are reported in Table \ref{tab:rho2}.

The newly determined $T_{1/2}$ of the $2^+_2$ state in $^{58}$Ni ($T_{1/2}$ = 0.60$^{+0.19}_{-0.12}$\,ps) is somewhat longer than, but consistent with, that reported previously in ($p,p'\gamma$) \cite{1969Be48}. 
The $T_{1/2}$ of the $2^+_2$ state in $^{60}$Ni ($T_{1/2}$ = 1.25$^{+0.76}_{-0.35}$\,ps) is inconsistent with that determined from the B($E2$) value from ($e,e'$) \cite{1983KL09}, but is in agreement with the limit determined from a previous ($n,n'\gamma$) study using reactor neutrons \cite{1989Ko54}.  The low abundance of $^{62}$Ni in the natural Ni scattering sample (3.63\,\%) prevented a measurement of the half-life of the 2$_2^+$ state; however, this $T_{1/2}$ was recently determined from the ($n,n'\gamma$) reaction at UK with an enriched scattering sample \cite{Chakraborty2011}.

The $\delta (E2/M1)$ mixing ratio of the 1321.2\,keV transition of $^{58}$Ni is measured to be -1.04$^{+0.07}_{-0.08}$ in the ($n,n'\gamma$) data. It was not possible to obtain a reliable result from our ($p,p'\gamma$) study for this transition. This new value is consistent with, but more precise than, the three previously reported values \cite{1967Ho,1969Be48,1971St02}, so it is adopted in this work.

For the 826.06\,keV \tra{2}{2}{2}{1} transition in $^{60}$Ni minima are found for $\delta$ at +0.40$^{+0.05}_{-0.04}$ or +1.01$^{+0.09}_{-0.10}$ from ($n,n'\gamma$). There is no consistency for this mixing ratio from previously reported values \cite{1965Mo12,Shafroth1966,1969Be48,VANPATTER1972355,Kearns1980,1989Ko54,2008To15}. In the ($p,p'\gamma$) data, we obtain a value of +0.63$^{+0.13}_{-0.10}$, which is consistent with one of the solutions from ($n,n'\gamma$).
We take the weighted mean of these two consistent values following the limitation of relative statistical weights (LRSW) procedure \cite{LRSW} to obtain the value reported in Table \ref{tab:rho2}.

In $^{62}$Ni for the 1128.82\,keV \tra{2}{2}{2}{1} transition, two solutions emerge from the ($p,p'\gamma$) data (+3.17(10) and -0.07(1)) and one is reported by Chakraborty {\it et al.} (+2.70$^{+0.38}_{-0.28}$) \cite{Chakraborty2011} from ($n,n'\gamma$) at UK. The other reported values again do not provide a consistent picture \cite{1969Be48,VANPATTER1972355,1976Ca31,1989Ko54}. We therefore again report the mean value of the new data in Table \ref{tab:rho2}.

Internal conversion coefficients measured in this work for pure $E2$ transitions match well with theoretical values. The $\rho^2 (E0)$ values for the \tra{0}{2}{0}{1} transitions in $^{60, 62}$Ni from internal conversion spectroscopy ($<$35, 132$^{+59}_{-70}$) in our work are consistent with values (1-27, 72$^{+60}_{-30}$) reported from detecting the internal pair formation (IPF) decays \cite{Passoja1981}.

\begin{table*}[!ht]
\centering
\caption{Pertinent experimental data for the determination of the $\rho^2 (E0)$ values for the \tra{2}{2}{2}{1} transitions of the $^{58,60,62}$Ni isotopes. The half-lives ($T_{1/2}$), branching ratios ($BR$), $\delta (E2/M1)$ mixing ratios, and internal conversion coefficients ($\alpha_K$) were obtained in this work with the exception of the $T_{1/2}$ and $BR$ for $^{62}$Ni, which were taken from Ref. \cite{Chakraborty2011}. 
}
\label{tab:rho2}
\begin{tabular}{cccccccccc}
\hline
 & E$_\gamma$ & $T_{1/2}$ & $BR$ & $\delta(E2/M1)$ & $\alpha_K \times 10^4$ & B($M1$) & B($E2$) & $M(E0)$ & $\rho^2(E0) \times 10^3$ \\
 & keV & ps & & & & $\mu_N^2$ & $e^2 fm^4$ & fm$^2$ & \\
\hline
\rule{0pt}{3ex} $^{58}$Ni & 1321.2 & 0.60$^{+0.19}_{-0.12}$ & 0.953(2) & -1.04$^{+0.07}_{-0.08}$ & 1.4(3) & 0.014(4) & 120$^{+30}_{-40}$ & 10.3$^{+1.1}_{-2.0}$ & 230$^{+50}_{-80}$ \\
\rule{0pt}{3ex} $^{60}$Ni & 826.06 & 1.28$^{+0.74}_{-0.35}$ & 0.860(5) & +0.43(8) & 3.0(1) & 0.046$^{+0.016}_{-0.021}$ & 195$^{+70}_{-110}$ & 9$^{+2}_{-4}$ & 150$^{+90}_{-110}$ \\
\rule{0pt}{3ex} $^{62}$Ni & 1128.82 & 0.67$^{+0.20}_{-0.14}$ & 0.45(4) & +3.1(1) & 2.0(1) & 1.8$^{+0.3}_{-0.5} \times 10^{-3}$ & 200$^{+40}_{-50}$ & 8.4$^{+1.4}_{-2.5}$ & 140$^{+50}_{-70}$ \\ 
\hline
\end{tabular}
\end{table*}

The new data for the $T_{1/2}$, $BR$, $\delta (E2/M1)$ and $\alpha_K$ are used in the calculation of $\rho^2 (E0)$, B($M1$) and B($E2$) values shown in Table \ref{tab:rho2} and Fig. \ref{fig:summary_results}.  The black (filled) arrows represent B($m\lambda$) values and the yellow (open) arrows show $\rho^2 (E0) \times 10^3$ values. The width of the arrow represents the strength of that component. In order to properly account for the asymmetric uncertainties in the various input quantities, these final values and uncertainties were calculated using a Monte Carlo method where all inputs are treated as a probability distribution. In the case where the final probability distribution returned regions of unphysical values, the Neyman construction \cite{Neyman} using the Feldman-Cousins ordering principle \cite{Feldman} was followed. As a consequence of using the Feldman-Cousins ordering principle, the median was selected to be the central value reported. The uncertainties in the $\rho^2 (E0)$ values are dominated by the statistical uncertainty in the internal conversion coefficients, measured here for the first time, and the precision in the half lives.

The other data shown in Fig. \ref{fig:summary_results} are the B($E2 ; 0_1^+ \rightarrow 2_1^+$) values taken from Allmond {\it et al.} \cite{Allmond.90.034309} determined recently by Coulomb excitation, the magnetic moments taken from Kenn {\it et al.} \cite{Kenn.63.021302} and the spectroscopic quadrupole moments from the Nuclear Data Sheets \cite{NucData58,NucData60,NucData62}.

\begin{figure}[!ht]
	\centering
	\includegraphics[width=1.0\linewidth]{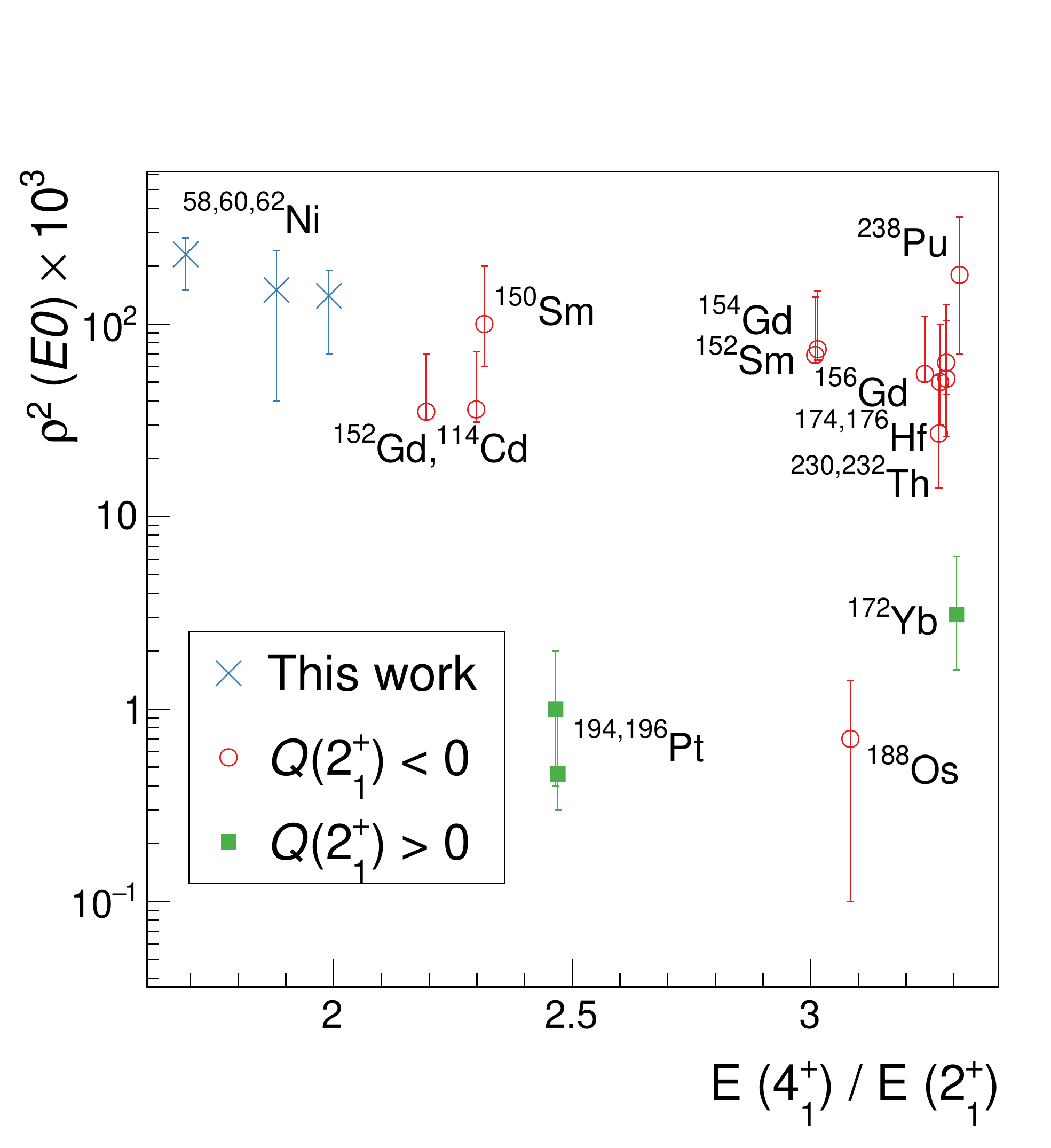}
	\caption{An overview of the experimentally measured $\rho^2 (E0) \times 10^3$ \tra{2}{2}{2}{1} values for the Ni isotopes from this work (cross) compared with the literature values taken from Ref. \cite{Wood1999} plotted vs. the energy ratio of the first 4$^+$ and 2$^+$ states. The literature values are grouped by the sign of the spectroscopic quadrupole moment of the 2$^+_1$ state, where negative (open circle) indicates prolate and positive (square) corresponds to oblate shape.
    }
\label{fig:ratio}
\end{figure}

These experiments represent the first measurements of the \tra{2}{2}{2}{1} $E0$ transitions in the Ni isotopes and in nuclei with spherical ground states. They are amongst the largest values reported across the chart of the nuclides. These new data are compared in Fig. \ref{fig:ratio} to the other \tra{2}{2}{2}{1} $E0$ transition strengths reported in the most recent survey of Wood {\it et al.} \cite{Wood1999}.

\section{Calculations and discussion\label{sect:shell}}
In a two-state mixing model, large $E0$ transition strengths are generated either by a large difference in mean-squared charge radii, or by a significant amount of mixing between the two states, or both.

The ground states of $^{58,60,62}$Ni have been determined to be spherical with little variation between them from detailed muonic X-ray measurements \cite{Shera1976} and optical spectroscopy \cite{Steudel1980}.
The spectroscopic quadrupole moment of the first 2$^+$ state in each isotope (-10(6), +3(5), +5(12)\,$efm^2$) is small \cite{Raghavan1989}, indicating that the first excited state is also close to spherical. In $^{58, 60}$Ni, the second 2$^+$ state lies below the second 0$^+$ state, thus excluding the possibility that these are members of co-existing deformed $K=0$ bands. From the systematic trend, it seems likely that the second 2$^+$ state in $^{62}$Ni also has the same structural origin as in the lighter isotopes. Another possibility is that these 2$^+$ states are the band-heads of $K^\pi = 2^+$ bands, however, a $\Delta K = 2$, $E0$ transition is forbidden in the axially symmetric rigid rotor model \cite{Wood1999}. In the spherical-quadrupole-vibrator model, $E0$ transitions are forbidden if the change in phonon number is one \cite{Wood1999}. Clearly, the predictions of these simple models do not explain the new $E0$ transition strength data.

A microscopic approach for the calculation of $E0$ transition strengths was recently described by Brown {\it et al.} \cite{Brown.95.011301}.  This approach takes the orbital occupations and one-body transition densities obtained from a configuration-interaction (CI) shell model calculation and uses them to constrain an energy density functional (EDF) calculation of the transition density. The resulting transition density contains valence terms coming from the model space, and the monopole polarization of the core protons by the valence nucleons. This first-order correction involves coupling of the valence orbitals to one-particle one-hole configurations of the giant monopole (breathing mode) excitation of the nucleus. The core polarization is determined by the shape of the valence transition density. The transition density is large only when there is a cancellation between the valence orbits that have a different number of nodes in the radial wavefunction. This method was applied to excited 0$^{+}$ to 0$^{+}$ ground-state transitions over a wide range of nuclei \cite{Brown.95.011301}. Overall, the theoretical $E0$ transition matrix element, $| M(E0)|$, values were a factor of 2 to 3 smaller than experiment. It was suggested that the remaining strength comes from second-order correlations involving collective 2$^{+}$ intermediate states \cite{Brown.95.011301}. The model is applicable to $E0$ transitions between states of any $J$ value and, in this work, we have used it to calculate $E0$ transition strengths between $2^+$ states for the first time.

The CI shell model calculations were performed for $^{58,60,62}$Ni with the NuShellX@MSU code \cite{NushellX}.  In each case, the calculations were performed in the full $fp$ model space using the GPFX1A interaction \cite{Honma2005}.

For the Ni isotopes, the valence wavefunctions are dominated
by neutron configurations that do not contribute to the $E0$ matrix element.
The valence orbits are $1p$ and $0f$, and the $E0$ transition
density comes from a cancellation between these two types of
radial wavefunction.  Ref. \cite{Brown.95.011301} considered
the 0$^{+}_{2} \rightarrow 0^{+}_{1}$ and 0$^{+}_{3} \rightarrow 0^{+}_{1}$ transitions in $^{58}$Ni. 
The calculated $M(E0)$ for the 0$^{+}_{2} \rightarrow 0^{+}_{1}$ transitions was much larger than the experimental value.
A significant improvement in the agreement was achieved through a remixing of the 0$^{+}_{2}$ - 0$^{+}_{3}$ and 2$^{+}_{2}$ - 2$^{+}_{3}$ states. The calculated $M(E0)$ for the remixed 0$^{+}$ states was about a factor of two smaller than experiment (comparable to the level of agreement achieved in the other nuclei studied in Ref. \cite{Brown.95.011301}).

\begin{table*}[ht]
\centering
\caption{Experimental and calculated properties of the first and second $2^+$ states in $^{58,60,62}$Ni. The calculated $M(E0)$ and $\rho^2 (E0) \times 10^3$ values are shown for the two Skyrme parameter sets $Skx$ and $s3$ \cite{Brown.95.011301}, see text for details.}
\label{tab:Theory-Exp}
\begin{tabular}{ccccccccccc}
\hline
& $E_2$ & $E_1$ & $\mu$ & $Q$ & B($M1$) & B($E2$) & \multicolumn{2}{c}{$M(E0)$} & \multicolumn{2}{c}{$\rho^2 (E0) \times 10^3$}\\
& $2_2^+$ & $2_1^+$ & $2_1^+$ & $2_1^+$ & \tra{2}{2}{2}{1} & \tra{2}{2}{2}{1} & & & & \\
& MeV & MeV & $\mu_N$ & $efm^2$ & $\mu_N^2$ & $e^2fm^4$ & \multicolumn{2}{c}{$fm^2$} & & \\
\hline
\multicolumn{11}{l}{Experiment:}\\
$^{58}$Ni & 2.77 & 1.45 & 0.076(18) & -10(6) & 0.014(4) & 120$^{+30}_{-40}$ & \multicolumn{2}{c}{10.3$^{+1.1}_{-2.0}$} & \multicolumn{2}{c}{230$^{+50}_{-80}$} \\
\rule{0pt}{3ex}$^{60}$Ni & 2.16 & 1.33 & 0.32(6) & 3(5) & 0.046$^{+0.016}_{-0.021}$ & 195$^{+70}_{-110}$ & \multicolumn{2}{c}{9$^{+2}_{-4}$} & \multicolumn{2}{c}{150$^{+90}_{-110}$} \\
\rule{0pt}{3ex}$^{62}$Ni & 2.30 & 1.17 & 0.33(5) & 5(12) & 1.8$^{+0.3}_{-0.5} \times 10^{-3}$ & 200$^{+40}_{-50}$ & \multicolumn{2}{c}{8.4$^{+1.4}_{-2.5}$} & \multicolumn{2}{c}{140$^{+50}_{-70}$} \\
\hline
\multicolumn{7}{l}{Theory:}& $Skx$ & $s3$ & $Skx$ & $s3$ \\
$^{58}$Ni & 2.64 & 1.48 & -0.14 & -2.7 & 0.165 & 30.3 & 1.66 & 0.80 & 5.9 & 1.4 \\
$^{60}$Ni & 2.29 & 1.56 & 0.30 & 2.3 & 0.044 & 269 & 0.38 & 0.50 & 0.3 & 0.5\\
$^{62}$Ni & 2.34 & 1.15 & 0.61 & 25.3 & 0.0039 & 151 & 0.53 & 1.06 & 0.6 & 2.2 \\
\hline
\end{tabular}
\end{table*}
The properties of the 2$^{+}_{2}$ and 2$^{+}_{1}$ states for $^{58,60,62}$Ni are compared in Table \ref{tab:Theory-Exp}. For the electromagnetic transitions, we use the effective $M1$ and $E2$ operators from Honma {\it et al.} \cite{Honma2004}. The calculated B($M1$ ; \tra{2}{2}{2}{1})  in $^{58}$Ni is an order of magnitude larger than experiment, and the other calculated B($M1$) and B($E2$) values are in reasonable agreement with experiment given the large error bars. The 2$^{+}_{1}$ state magnetic ($\mu$) and quadrupole ($Q$) matrix elements are also in reasonable agreement between theory and experiment.

The EDF part of the calculation used the $Skx$ and $s3$ Skyrme parameter sets
that were used in Ref. \cite{Brown.95.011301}.
The calculated values of $| M(E0)|$ are significantly smaller than experiment.
The value of $M(E0)$ obtained from the calculation in Table \ref{tab:Theory-Exp} is the maximum achievable with the current method.
For example, for $^{58}$Ni the wavefunctions are dominated by the two-neutron configurations outside of a $^{56}$Ni closed core. The only possible two-neutron configurations for 2$^{+}$ are $[(1p_{3/2})^{2}]$, $[(0f_{5/2})^{2}]$, $[1p_{1/2},0f_{5/2}]$, and $[1p_{3/2},0f_{5/2}]$.
If only pure configurations are considered, all of the off-diagonal $E0$ matrix elements are zero. A non-zero matrix element comes from only mixed configurations (as in the example of the $1p_{1/2}$ and $0g_{9/2}$ orbitals in $^{90}$Zr \cite{Brown.95.011301}).

We also consider the $jj44$ model space which allows for neutrons in the $(0f_{5/2},1p_{3/2},1p_{1/2},0g_{9/2})$ orbitals. With the $jj44a$ interaction of Lisetskiy {\it et al.} \cite{Lisetskiy2004}, the occupancy of the $0g_{9/2}$ orbit is small for $^{58,60,62}$Ni. These calculated neutron occupation numbers agree well with those determined in transfer reactions \cite{Schiffer2012,Schiffer2013}. The values of $M(E0)$ using the $jj44a$ and $pf$ model spaces are similar and substantially smaller than experiment.

\sloppy
Reproducing the experimental values of $M(E0)$, which are significantly larger than the present calculation may require second-order corrections involving the coupling of the valence nucleons to the one-particle one-hole giant quadrupole excitation of the nucleus. This effect is related to the increase in the mean-squared charge radii due to deformation or zero-point quadrupole oscillations \cite{Uher1966}, which involve the diagonal $E0$ matrix elements. Similar second-order corrections should be considered for the off-diagonal matrix elements involved in the $E0$ decays. To get an estimate for the magnitude of the second-order correction, we consider the data for the difference in mean-squared radii for $^{56}$Fe and $^{54}$Fe \cite{Minamisono2016} which, for the total matrix element is multiplied by the atomic number, is $Z \delta\langle r^{2}\rangle$ = 8.2(2) fm$^{2}$.  
Our EDF calculations give 2.6 fm$^{2}$ with $Skx$ and 2.1 fm$^{2}$ with $s3$ (similar to the EDF calculations shown in Fig. 2 of Ref. \cite{Minamisono2016}). The difference between the experimental mean-squared charge radii and the EDF calculations is similar in magnitude to the difference we observe for the $E0$ transition strengths.
Quantitative calculations for second-order corrections to the isotopic change and $E0$ matrix elements remain to be carried out and could be rather substantive.

\section{Conclusion}
Gamma-ray and electron spectroscopy measurements performed at the Australian National University and the University of Kentucky have been combined to measure the $E0$ transition strengths for the \tra{2}{2}{2}{1} transitions of $^{58, 60, 62}$Ni. The $E0$ strengths are found to be unexpectedly large.  Reported in this work are the first measurements of $E0$ transition strengths between 2$^+$ states in nuclei with spherical ground states. The microscopic model of Brown {\it et al.} \cite{Brown.95.011301} has been applied here for the first time for transitions between 2$^+$ states and, although this model is successful in reproducing $E0$ transition strengths in \tra{0}{}{0}{} cases, it does not reproduce the new experimental results. The origin of the large \tra{2}{}{2}{} $\rho^2 (E0)$ values in these isotopes should be the focus of future developments and refinement to theoretical models. Second-order corrections to the microscopic model involving the coupling of the valence nucleons to the one-particle one-hole giant quadrupole excitation of the nucleus are one development for future consideration.

\section*{Acknowledgments}
\sloppy
We would like to thank the technical staff of the Heavy Ion Accelerator Facility at the Australian National University, and in particular Justin Heighway for preparing the nickel targets. A.B.G. is grateful for support from the Department of Nuclear Physics of the Australian National University. This work was supported in part by the Natural Sciences and Engineering Research Council of Canada (NSERC); the National Science Foundation, Grants No. PHY-1404442 and PHY-1606890; and by the Australian Research Council Discovery Grants DP140102986 and FT100100991. TRIUMF receives funding via a contribution agreement through the National Research Council Canada.

\section*{References}

\bibliography{main}

\end{document}